\documentclass[aps,twocolumn,groupedaddress,altaffilletter]{revtex4}
\usepackage[dvips]{graphicx,color}
\usepackage{times}
\makeatletter
\def\frontmatter@thefootnote{%
   \altaffilletter@sw{\@alph}{\@alph}\c@footnote}
\def\frontmatter@makefnmark{%
   \@textsuperscript{\normalfont\@thefnmark)}}
\makeatother

\begin{document}


\title{Self-assembly of laterally aligned GaAs quantum dot pairs}


\author{M. Yamagiwa}
\email[Author to whom correspondence should be addressed; electronic mail:]{YAMAGIWA.Masakazu@nims.go.jp}
\affiliation{Quantum Dot Research Center, National Institute for Materials Science, 1-2-1 Sengen,
Tsukuba 305-0047, Japan.}
\author{T. Mano}
\affiliation{Quantum Dot Research Center, National Institute for Materials Science, 1-2-1 Sengen,
Tsukuba 305-0047, Japan.}
\author{T. Kuroda}
\altaffiliation[Also at: ]{PRESTO, Japan Science and Technology Agency.}
\affiliation{Quantum Dot Research Center, National Institute for Materials Science, 1-2-1 Sengen,
Tsukuba 305-0047, Japan.}
\author{T. Tateno}
\author{K. Sakoda}
\author{G. Kido}
\author{N. Koguchi}
\affiliation{Quantum Dot Research Center, National Institute for Materials Science, 1-2-1 Sengen,
Tsukuba 305-0047, Japan.}
\author{F. Minami}
\affiliation{Department of Physics, Tokyo Institute of Technology, Meguro-ku, Tokyo 152-8551, Japan.}

\date{\today}

\begin{abstract}
We report the fabrication of self-assembled, strain-free GaAs/Al$_{0.27}$Ga$_{0.73}$As quantum dot pairs which are laterally aligned in the growth plane, utilizing the droplet epitaxy technique and the anisotropic surface potentials of the GaAs (100) surface for the migration of Ga adatoms.  Photoluminescence spectra from a single quantum dot pair, consisting of a doublet, have been observed.  Finite element energy level calculations of a model quantum dot pair are also presented.

\end{abstract}


\maketitle


When two semiconductor quantum dots (QDs), which can each spatially confine an individual carrier in a discrete energy level, are in close proximity to each other, these carriers begin to interact with each other.  Specifically, the wavefunctions of the carriers confined in each QD of the QD pair (QDP) begin to overlap, resulting in an efficient carrier tunneling \cite{sche97,baye01}, and the wavefunctions may become admixed to form molecular orbital states.  Moreover, resonance in the optical transition energies leads to the formation of a coupled QDP via dipole-dipole interactions \cite{unol05,yama05,st06}.  Proposals for using such a coupling in quantum information processing have been brought forth \cite{biol00,he05}.  To this end, various semiconductor QDPs have been fabricated and studied, such as coupled QDs grown by cleaved edge quantum well overgrowth \cite{sche97}, vertically--aligned QDs grown by Stranski-Krastanow  
epitaxy incorporating an indium-flush procedure \cite{baye01,yama05}, and interface QDs formed in a quantum well \cite{unol05}. 
In this work we demonstrate the fabrication of \textit{self-assembled}, 
\textit{laterally aligned} GaAs/Al$_{0.27}$Ga$_{0.73}$As QDP structures by droplet epitaxy: Droplet epitaxy is a nonconventional growth technique 
for the self-assembly of high quality nanostructures 
with lattice-matched materials, making possible the growth of 
nanostructures with various structural characteristics, such as 
QDs \cite{kogu91,kogu93,wata00}, quantum rings (QRs) \cite{mano05a}, and concentric double QRs 
\cite{mano05b,kuro05}.
Here we observe the formation of laterally aligned GaAs QDPs by accurately selecting the As$_{4}$ flux during crystallization. Within relevant conditions, the nanocrystals show a remarkable shape like a double summit, reflecting the anisotropic surface potentials of the GaAs (100) surface.  To gain insight into the interactions between the carriers confined in such a structure, we study the QDP single-structure optical properties.  We further consider the electronic structure of a model QDP via energy level calculations requiring no inclusion of strain.

\begin{figure}
	\centering
		\includegraphics[scale=0.40]{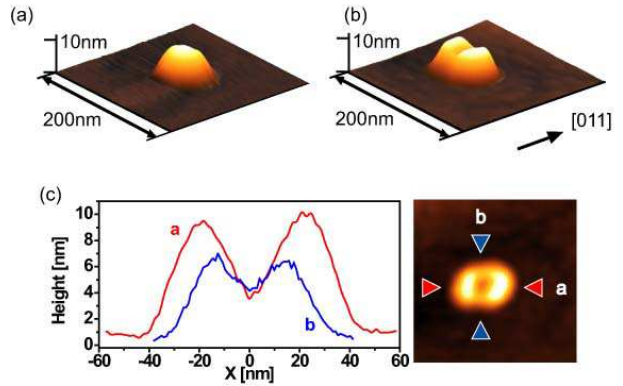}
	\caption{(color online) Atomic force microscope images of a Ga droplet (a), and a GaAs QDP (b).  These images are $200 \times 200$ nm$^{2}$ in size, and the dark-to-light contrast is 9 nm.  The line profile taken in the [0$\bar{1}$1] direction (labeled by a) and the [011] direction (labeled by b) are shown in (c).}
	\label{fig:fig1}
\end{figure}

The samples are grown by droplet epitaxy using a molecular beam epitaxy 
system with elemental sources and a valved As source, which enables the 
accurate control of the As$_{4}$ flux.  The layered epitaxial materials are grown on a GaAs (100) substrate.   
First, Ga droplets are formed by the following conditions: 1.5 monolayers (MLs) of Ga are supplied to a c($4 \times 4$) surface with a flux of 0.05~ML/s, and a substrate temperature of 330$^{\circ}$C \cite{yama04}.  The Ga droplets are nearly hemispherical in shape [Fig. 1(a)].  The average base size and height estimated from atomic force microscopy (AFM) measurements
are 63 ($\pm$ 5) nm and 9 ($\pm$ 1) nm, respectively.   Here, the density of the 
droplets (and consequently that of the GaAs crystals) is $2 \times 10^{8}$ cm$^{-2}$;  a low density is desired in order to obtain the photoluminescence (PL) from a single structure. 
The supplied Ga is less than the well known As coverage of a c($4 \times 4$) surface (1.75ML)~\cite{falt92}; yet, Ga droplet formation with 1.5 MLs of Ga is possible, depending on the surface reconstruction (the As coverage is 1.0 ML for a c(4$\times$4)-$\alpha$ surface \cite{ohta04}).
Next, the Ga droplets are crystallized into GaAs at 200$^{\circ}$C by the irradiation of an As$_{4}$ flux of $4 \times 10^{-5}$ Torr beam equivalent pressure. 
In this case, formation of GaAs nanocrystals clearly split in the [0$\bar{1}$1] direction is observed [Fig. 1(b)]. The cross-sectional image shows that the peaks in the [0$\bar{1}$1] direction [line a in Fig. 1(c)] are clearly higher than the peaks in the [011] direction [line b], with the peak-to-valley differences being more distinct in the [0$\bar{1}$1] direction.  This structure basically consists of two QDs aligned in the [0$\bar{1}$1] direction.  The average base size and height of each QD is 45 ($\pm$ 3) nm and 10 ($\pm$ 2) nm, respectively, and are separated by an average distance of 39 ($\pm$ 2) nm between their apexes, as estimated from AFM measurements. For the study of its optical properties, the GaAs QDPs are embedded in an Al$_{0.27}$Ga$_{0.73}$As barrier layer, and annealed by rapid thermal annealing after the entire growth \cite{mano05b}. 

The formation of this QDP can be understood by a crystallization mechanism which involves two processes.  These are the droplet-edge enhanced crystallization (Process A), and the crystallization which occurs outside of the droplet edges, which is induced by the surface counter migrations of Ga away from the droplet and As toward the droplet (Process B) \cite{mano05c}.  Process A is relatively isotropic due to the hemispherical shape of the droplet.  Process B is anisotropic, due to the anisotropic surface potentials of the GaAs (001) surface for the migration of Ga adatoms, and is strongly affected by the crystallization condition.  In this case, in the [011] direction, the Ga adatoms do not migrate far from the droplet, so Processes A and B occur near the edge of the droplet.  In the [0$\bar{1}$1] direction, the Ga adatoms migrate further from the droplet, so Process A occurs at the droplet edge while Process B occurs at a distance from the edge of the droplet.  This results in the structural anisotropy. 
The details of this mechanism will be addressed in future works.

\begin{figure}[htbp]
		\includegraphics[scale=0.40]{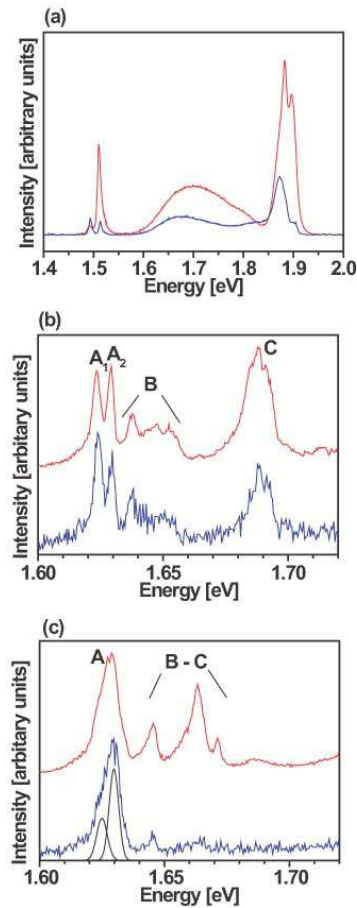}
	\label{fig:fig2}
	\caption{(color online) (a) The PL spectra of QDP ensembles.  The spectrum with lower intensity is excited at 7.6 W/cm$^{2}$, the one with higher intensity at 38 W/cm$^{2}$.  (b),(c) Two examples of the $\mu$PL spectra from a single QDP.  The excitation densities for each QDP are, from bottom to top, 2.1 W/cm$^{2}$ and 21 W/cm$^{2}$.  The spectra are normalized to their maxima and offset for clarity.  The black lines in (c) are Gaussian fits of two peaks to A.}
\end{figure}

A study of the optical properties of QDPs gives insight into their confined energy levels, and will be discussed in detail.  The PL spectra of QDP ensembles
at 4.2 K, using an Ar$^{+}$ laser (488 nm) as an excitation source, 
are shown in Fig. 2(a).
At low excitation (7.6 W/cm$^{2}$), the spectrum shows a 
broad peak centered at 1.67 eV, with a full width at half maximum  
of 80 meV.  As the excitation density is increased to 38 W/cm$^{2}$, a blueshift of the 
PL spectrum is seen, indicating the appearance of carrier recombinations 
of higher energy states, caused by state filling.  The peaks at $\sim$1.5 eV originates from the GaAs substrate, and those at $\sim$1.9 eV originate from the bulk Al$_{0.27}$Ga$_{0.73}$As and/or a two dimensional underlying GaAs layer \cite{kogu93}.

In the micro-PL ($\mu$PL) measurements of single QDPs, the sample is excited at 4.2 K using a HeNe laser emitting at 544 nm.  The excitation beam is focused onto the sample by a 50$\times$ objective lens. The PL is picked up by the same objective, dispersed by a polychromator with a 32 cm focal length, and recorded with a charge coupled device camera. 
The position of the excitation/observation 
spot on the sample is line-scanned with sub-micrometer precision. 
As the excitation
spot is translated, the spectra show a group of peaks which appear and then disappear in unison.  This spatial correlation allows one to determine the spectral peaks which originate from a single QDP. 

The $\mu$PL spectrum of two different, single QDPs -- \textit{a} and \textit{b} -- are shown in Figs. 2(b) and 2(c), respectively. 
In 
QDP-\textit{a}, two luminescent lines appear at approximately 1.623 eV (A$_{1}$), and 1.629 eV (A$_{2}$), together with several peaks centered at 1.646eV (B), and 1.688eV (C).
At high excitation, C becomes more pronounced.  A similar behavior can be observed in 
QDP-\textit{b}, 
with an anti-symmetric peak at 1.628 eV (A), which may consist of a doublet with $\sim$4.8 meV spacing, and several peaks at 1.628--1.663 eV (B--C). These PL peaks may
originate from various states in a single QDP.  The spectral characteristics of this QDP are different from those of a GaAs/AlGaAs QD \cite{yama04} or QR \cite{kuro05}.  Specifically, the spacings between the doublets are much larger than the biexciton binding energy (0.95 meV) determined from GaAs/AlGaAs QDs \cite{kuro06}, indicating that these doublets do not originate from multi-excitonic emissions.

\begin{figure}[htbp]
	\centering
		\includegraphics[scale=0.55]{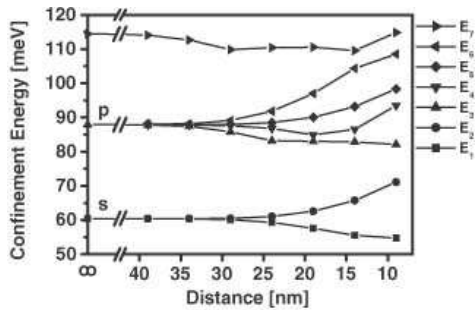}
	\caption{The calculated first seven eigenvalues (E$_{1}$-E$_{7}$) of the confined electronic states in a model QDP as a function of the peak-to-peak distance between the two QDs.}
	\label{fig:fig3}
\end{figure}

The energy levels for a structure similar to the QDP can be calculated in a fairly straightforward manner, due to the lack of strain in the system.  In order to obtain insight into the origins of these emission lines from a single QDP, the electronic energy levels of a model QDP
and their wavefunctions are calculated in the framework of finite element methods with strain-free, single-band effective mass approximations. For simplicity, the QDP is assumed to be two truncated, overlapping cones -- QDs -- which are symmetric and are of a base diameter of 44 nm and a height of 10 nm; the sizes are equivalent to those of the QDP. The conduction band effective mass is 0.067$m$$_{0}$ in GaAs and 0.089$m$$_{0}$ in Al$_{0.27}$Ga$_{0.73}$As\cite{pave94}, and the  
band offset is 220 meV\cite{bosi88,pave94,yama04}. (Note that our calculation does not include the confinement of holes, because their contributions to the spectral spacings are one order of magnitude smaller than those of electrons.)
The first seven eigenvalues of the confined electronic states, $E_{1}$--$E_{7}$, are shown in Fig. 3 as a function of the peak-to-peak distance between the two QDs.  As the QDs are brought closer together and begin to overlap, single QD levels are admixed, and molecular-orbital energy levels appear.  The 
first two eigenstates, 
$E_{1}$ and $E_{2}$, which correspond to the \textit{s}-like single-QD wavefunctions, reveal the formation of bonding/antibonding states ($s\sigma_{g}$ and $s\sigma_{u}$), which spread apart with decreasing distance.  $E_{3}$, $E_{4}$, $E_{5}$ and $E_{6}$ form the \textit{p} states ($p\sigma_{g}$, $p\pi_{g}$, $p\pi_{u}$ and $p\sigma_{u}$), which also spread apart with decreasing distance. If one compares these 
energy level spacings to the $\mu$PL emission line spacings of QDP-\textit{a} and -\textit{b},
one finds similarities between the two, i.e. the doublet A emissions has a spacing close to that of the molecular states formed by the \textit{s}-like QD state, and B to those of the \textit{p} states.  In this case, the differences in the $\mu$PL emission line spacings for QDP-\textit{a} and -\textit{b} can be interpreted here as being due to the different distance between the QDs, the distance being smaller for QDP-\textit{a} than for QDP-\textit{b} (i.e., in Fig. 2(b), A$_{1}$ and A$_{2}$, as well as the peaks in B, are spread further apart than the corresponding peaks in Fig. 2(c)).

Although the above calculations seem to indicate a tunnel coupling between the QDs, the origin of the emission lines from a single QDP are still open to debate.  It should be noted here that the asymmetry of the QDP is expected to be not as enhanced as would be the case of, e.g., stacked QDs, due to the difference in the microscopic pair--formation mechanism.  In the growth of the QDP, a pair of QDs originates from a single Ga droplet, with its crystalline shape being changed from a hemisphere to a double summit according to the uniaxial diffusion of Ga atoms.  Since the atomic diffusion is not of a polar process but of an axial one (i.e., the efficiency of diffusion is identical along the x and $-$x direction), it is expected that the shape of the QDP has a reversal symmetry, and that it consists of identically--formed QDs.  The situation is quite different for that of stacked QDs, in which the two QDs are independently formed in a stochastic manner.  However, even in the QDP formation, microscopic effects which would perturb the QDP symmetry cannot be ruled out, and AFM measurements alone do not guarantee the lack of a large asymmetry.  Thus, to properly assign the origins of the emission lines from a single QDP, future work include the measurement of a spectral peak shift with an applied, in-plane electric field, photon-correlation measurements, as well as time resolved spectral evolution studies.  

To conclude, the fabrication of a strain-free, laterally aligned GaAs/Al$_{0.27}$Ga$_{0.73}$As QDP structure using droplet epitaxy is demonstrated by utilizing the anisotropic surface potentials of the GaAs (100) surface, the effects of which are induced by the selection of the appropriate As$_{4}$ flux.  
The $\mu$PL spectra of a single QDP shows the ensemble of emissions which, according to simple calculations of the electronic energy levels of a model QDP, may indicate the existence of a tunnel coupling between the QDs.  However, it must be stated that the origins of these emission lines are still not fully understood.  Further characterization is necessary to determine the origins of the emission lines, and to establish the electronic structure of QDPs.

We are grateful to Drs. J. S. Kim, T. Noda, A. Ohtake, Prof. S. Sanguinetti, and Prof.M. Kawabe for fruitful discussions.  This work was partially supported by the MEXT.


\begin{thebibliography}{20}
\expandafter\ifx\csname natexlab\endcsname\relax\def\natexlab#1{#1}\fi
\expandafter\ifx\csname bibnamefont\endcsname\relax
  \def\bibnamefont#1{#1}\fi
\expandafter\ifx\csname bibfnamefont\endcsname\relax
  \def\bibfnamefont#1{#1}\fi
\expandafter\ifx\csname citenamefont\endcsname\relax
  \def\citenamefont#1{#1}\fi
\expandafter\ifx\csname url\endcsname\relax
  \def\url#1{\texttt{#1}}\fi
\expandafter\ifx\csname urlprefix\endcsname\relax\def\urlprefix{URL }\fi
\providecommand{\bibinfo}[2]{#2}
\providecommand{\eprint}[2][]{\url{#2}}

\bibitem[{\citenamefont{Schedelbeck et~al.}(1997)\citenamefont{Schedelbeck,
  Wegscheider, Bichler, and Abstreiter}}]{sche97}
\bibinfo{author}{\bibfnamefont{G.}~\bibnamefont{Schedelbeck}},
  \bibinfo{author}{\bibfnamefont{W.}~\bibnamefont{Wegscheider}},
  \bibinfo{author}{\bibfnamefont{M.}~\bibnamefont{Bichler}}, \bibnamefont{and}
  \bibinfo{author}{\bibfnamefont{G.}~\bibnamefont{Abstreiter}},
  \bibinfo{journal}{Science} \textbf{\bibinfo{volume}{278}},
  \bibinfo{pages}{1792} (\bibinfo{year}{1997}).

\bibitem[{\citenamefont{Bayer et~al.}(2001)\citenamefont{Bayer, Hawrylak,
  Hinzer, Fafard, Korkusinki, Wasilewski, Stern, and Forchel}}]{baye01}
\bibinfo{author}{\bibfnamefont{M.}~\bibnamefont{Bayer}},
  \bibinfo{author}{\bibfnamefont{P.}~\bibnamefont{Hawrylak}},
  \bibinfo{author}{\bibfnamefont{K.}~\bibnamefont{Hinzer}},
  \bibinfo{author}{\bibfnamefont{S.}~\bibnamefont{Fafard}},
  \bibinfo{author}{\bibfnamefont{M.}~\bibnamefont{Korkusinki}},
  \bibinfo{author}{\bibfnamefont{Z.~R.} \bibnamefont{Wasilewski}},
  \bibinfo{author}{\bibfnamefont{O.}~\bibnamefont{Stern}}, \bibnamefont{and}
  \bibinfo{author}{\bibfnamefont{A.}~\bibnamefont{Forchel}},
  \bibinfo{journal}{Science} \textbf{\bibinfo{volume}{291}},
  \bibinfo{pages}{451} (\bibinfo{year}{2001}).
  
\bibitem[{\citenamefont{Unold et~al.}(2005)\citenamefont{Unold, Mueller,
  Lienau, Elsaesser, and Wieck}}]{unol05}
\bibinfo{author}{\bibfnamefont{T.}~\bibnamefont{Unold}},
  \bibinfo{author}{\bibfnamefont{K.}~\bibnamefont{Mueller}},
  \bibinfo{author}{\bibfnamefont{C.}~\bibnamefont{Lienau}},
  \bibinfo{author}{\bibfnamefont{T.}~\bibnamefont{Elsaesser}},
  \bibnamefont{and} \bibinfo{author}{\bibfnamefont{A.~D.} \bibnamefont{Wieck}},
  \bibinfo{journal}{Phys.\ Rev. Lett.} \textbf{\bibinfo{volume}{94}},
  \bibinfo{pages}{137404} (\bibinfo{year}{2005}).

\bibitem[{\citenamefont{Yamauchi et~al.}(2005)\citenamefont{Yamauchi, Komori,
  Morohashi, Goshima, Sugaya, and Takagahara}}]{yama05}
\bibinfo{author}{\bibfnamefont{S.}~\bibnamefont{Yamauchi}},
  \bibinfo{author}{\bibfnamefont{K.}~\bibnamefont{Komori}},
  \bibinfo{author}{\bibfnamefont{I.}~\bibnamefont{Morohashi}},
  \bibinfo{author}{\bibfnamefont{K.}~\bibnamefont{Goshima}},
  \bibinfo{author}{\bibfnamefont{T.}~\bibnamefont{Sugaya}}, \bibnamefont{and}
  \bibinfo{author}{\bibfnamefont{T.}~\bibnamefont{Takagahara}},
  \bibinfo{journal}{Appl. Phys.\ Lett.} \textbf{\bibinfo{volume}{87}},
  \bibinfo{pages}{182103} (\bibinfo{year}{2005}).
      
\bibitem[{\citenamefont{Stinaff et~al.}(2006)\citenamefont{Stinaff, Scheibner, Bracker, Ponomarev, Korenev, Ware, Doty, Reinecke, and Gammon}}]{st06}
\bibinfo{author}{\bibfnamefont{E.~A.}~\bibnamefont{Stinaff}},
  \bibinfo{author}{\bibfnamefont{M.}~\bibnamefont{Scheibner}},
  \bibinfo{author}{\bibfnamefont{A.~S.}~\bibnamefont{Bracker}},
  \bibinfo{author}{\bibfnamefont{I.~V.}~\bibnamefont{Ponomarev}},
  \bibinfo{author}{\bibfnamefont{V.~L.}~\bibnamefont{Korenev}},
  \bibinfo{author}{\bibfnamefont{M.~E.}~\bibnamefont{Ware}},
  \bibinfo{author}{\bibfnamefont{M.~F.}~\bibnamefont{Doty}},
  \bibinfo{author}{\bibfnamefont{T.~L.}~\bibnamefont{Reinecke}},
   \bibnamefont{and}
   \bibinfo{author}{\bibfnamefont{D.}~\bibnamefont{Gammon}},
  \bibinfo{journal}{Science} \textbf{\bibinfo{volume}{311}},
  \bibinfo{pages}{636} (\bibinfo{year}{2006}).

\bibitem[{\citenamefont{Biolatti et~al.}(2000)\citenamefont{Biolatti, Iotti,
  Zanardi, and Rossi}}]{biol00}
\bibinfo{author}{\bibfnamefont{E.}~\bibnamefont{Biolatti}},
  \bibinfo{author}{\bibfnamefont{R.~C.} \bibnamefont{Iotti}},
  \bibinfo{author}{\bibfnamefont{P.}~\bibnamefont{Zanardi}}, \bibnamefont{and}
  \bibinfo{author}{\bibfnamefont{F.}~\bibnamefont{Rossi}},
  \bibinfo{journal}{Phys.\ Rev. Lett.} \textbf{\bibinfo{volume}{85}},
  \bibinfo{pages}{5647} (\bibinfo{year}{2000}).
  
\bibitem[{\citenamefont{He et~al.}(2005)\citenamefont{He, Bester, and
  Zunger}}]{he05}
\bibinfo{author}{\bibfnamefont{L.}~\bibnamefont{He}},
  \bibinfo{author}{\bibfnamefont{G.}~\bibnamefont{Bester}}, \bibnamefont{and}
  \bibinfo{author}{\bibfnamefont{A.}~\bibnamefont{Zunger}},
  \bibinfo{journal}{Phys.\ Rev. B} \textbf{\bibinfo{volume}{72}},
  \bibinfo{pages}{081311(R)} (\bibinfo{year}{2005}).
  
\bibitem[{\citenamefont{Koguchi et~al.}(1991)\citenamefont{Koguchi, Takahashi,
  and Chikyow}}]{kogu91}
\bibinfo{author}{\bibfnamefont{N.}~\bibnamefont{Koguchi}},
  \bibinfo{author}{\bibfnamefont{S.}~\bibnamefont{Takahashi}},
  \bibnamefont{and} \bibinfo{author}{\bibfnamefont{T.}~\bibnamefont{Chikyow}},
  \bibinfo{journal}{J.\ Cryst. Growth} \textbf{\bibinfo{volume}{111}},
  \bibinfo{pages}{688} (\bibinfo{year}{1991}).

\bibitem[{\citenamefont{Koguchi and Ishige}(1993)}]{kogu93}
\bibinfo{author}{\bibfnamefont{N.}~\bibnamefont{Koguchi}} \bibnamefont{and}
  \bibinfo{author}{\bibfnamefont{K.}~\bibnamefont{Ishige}},
  \bibinfo{journal}{Jpn.\ J. Appl. Phys.} \textbf{\bibinfo{volume}{32}},
  \bibinfo{pages}{2052} (\bibinfo{year}{1993}).

\bibitem[{\citenamefont{Watanabe et~al.}(2000)\citenamefont{Watanabe, Koguchi,
  and Gotoh}}]{wata00}
\bibinfo{author}{\bibfnamefont{K.}~\bibnamefont{Watanabe}},
  \bibinfo{author}{\bibfnamefont{N.}~\bibnamefont{Koguchi}}, \bibnamefont{and}
  \bibinfo{author}{\bibfnamefont{Y.}~\bibnamefont{Gotoh}},
  \bibinfo{journal}{Jpn.\ J. Appl. Phys.} \textbf{\bibinfo{volume}{39}},
  \bibinfo{pages}{L79} (\bibinfo{year}{2000}).

\bibitem[{\citenamefont{Mano and Koguchi}(2005)}]{mano05a}
\bibinfo{author}{\bibfnamefont{T.}~\bibnamefont{Mano}} \bibnamefont{and}
  \bibinfo{author}{\bibfnamefont{N.}~\bibnamefont{Koguchi}},
  \bibinfo{journal}{J.\ Cryst. Growth} \textbf{\bibinfo{volume}{278}},
  \bibinfo{pages}{108} (\bibinfo{year}{2005}).

\bibitem[{\citenamefont{Mano et~al.}(2005)\citenamefont{Mano, Kuroda,
  Sanguinetti, Ochiai, Tateno, Kim, Noda, Kawabe, Sakoda, Kido
  et~al.}}]{mano05b}
\bibinfo{author}{\bibfnamefont{T.}~\bibnamefont{Mano}},
  \bibinfo{author}{\bibfnamefont{T.}~\bibnamefont{Kuroda}},
  \bibinfo{author}{\bibfnamefont{S.}~\bibnamefont{Sanguinetti}},
  \bibinfo{author}{\bibfnamefont{T.}~\bibnamefont{Ochiai}},
  \bibinfo{author}{\bibfnamefont{T.}~\bibnamefont{Tateno}},
  \bibinfo{author}{\bibfnamefont{J.~S.} \bibnamefont{Kim}},
  \bibinfo{author}{\bibfnamefont{T.}~\bibnamefont{Noda}},
  \bibinfo{author}{\bibfnamefont{M.}~\bibnamefont{Kawabe}},
  \bibinfo{author}{\bibfnamefont{K.}~\bibnamefont{Sakoda}},
  \bibinfo{author}{\bibfnamefont{G.}~\bibnamefont{Kido}},  \bibnamefont{and}
  \bibinfo{author}{\bibfnamefont{N.}~\bibnamefont{Koguchi}}, 
  \bibinfo{journal}{Nano\ Letters} \textbf{\bibinfo{volume}{5}},
  \bibinfo{pages}{425} (\bibinfo{year}{2005}).

\bibitem[{\citenamefont{Kuroda et~al.}(2005)\citenamefont{Kuroda, Mano, Ochiai,
  Sanguinetti, Sakoda, Kido, and Koguchi}}]{kuro05}
\bibinfo{author}{\bibfnamefont{T.}~\bibnamefont{Kuroda}},
  \bibinfo{author}{\bibfnamefont{T.}~\bibnamefont{Mano}},
  \bibinfo{author}{\bibfnamefont{T.}~\bibnamefont{Ochiai}},
  \bibinfo{author}{\bibfnamefont{S.}~\bibnamefont{Sanguinetti}},
  \bibinfo{author}{\bibfnamefont{K.}~\bibnamefont{Sakoda}},
  \bibinfo{author}{\bibfnamefont{G.}~\bibnamefont{Kido}}, \bibnamefont{and}
  \bibinfo{author}{\bibfnamefont{N.}~\bibnamefont{Koguchi}},
  \bibinfo{journal}{Phys.\ Rev. B} \textbf{\bibinfo{volume}{72}},
  \bibinfo{pages}{205301} (\bibinfo{year}{2005}).

\bibitem[{\citenamefont{Yamagiwa et~al.}(2004)\citenamefont{Yamagiwa, Sumita,
  Minami, and Koguchi}}]{yama04}
\bibinfo{author}{\bibfnamefont{M.}~\bibnamefont{Yamagiwa}},
  \bibinfo{author}{\bibfnamefont{N.}~\bibnamefont{Sumita}},
  \bibinfo{author}{\bibfnamefont{F.}~\bibnamefont{Minami}}, \bibnamefont{and}
  \bibinfo{author}{\bibfnamefont{N.}~\bibnamefont{Koguchi}},
  \bibinfo{journal}{J.\ Lumin.} \textbf{\bibinfo{volume}{108}},
  \bibinfo{pages}{379} (\bibinfo{year}{2004}).

\bibitem[{\citenamefont{Falta et~al.}(1992)\citenamefont{Falta, Tromp, Copel,
  Pettit, and Kirchner}}]{falt92}
\bibinfo{author}{\bibfnamefont{J.}~\bibnamefont{Falta}},
  \bibinfo{author}{\bibfnamefont{R.~M.} \bibnamefont{Tromp}},
  \bibinfo{author}{\bibfnamefont{M.}~\bibnamefont{Copel}},
  \bibinfo{author}{\bibfnamefont{G.~D.} \bibnamefont{Pettit}},
  \bibnamefont{and} \bibinfo{author}{\bibfnamefont{P.~D.}
  \bibnamefont{Kirchner}}, \bibinfo{journal}{Phys.\ Rev. Lett.}
  \textbf{\bibinfo{volume}{69}}, \bibinfo{pages}{3068} (\bibinfo{year}{1992}).

\bibitem[{\citenamefont{Ohtake et~al.}(2004)\citenamefont{Ohtake, Kocan,
  Nakamura, Natori, and Koguchi}}]{ohta04}
\bibinfo{author}{\bibfnamefont{A.}~\bibnamefont{Ohtake}},
  \bibinfo{author}{\bibfnamefont{P.}~\bibnamefont{Kocan}},
  \bibinfo{author}{\bibfnamefont{J.}~\bibnamefont{Nakamura}},
  \bibinfo{author}{\bibfnamefont{A.}~\bibnamefont{Natori}}, \bibnamefont{and}
  \bibinfo{author}{\bibfnamefont{N.}~\bibnamefont{Koguchi}},
  \bibinfo{journal}{Phys.\ Rev. Lett.} \textbf{\bibinfo{volume}{92}},
  \bibinfo{pages}{236105} (\bibinfo{year}{2004}).

\bibitem[{\citenamefont{Mano et~al.}(2006)\citenamefont{Mano, Noda, Yamagiwa,
  and Koguchi}}]{mano05c}
\bibinfo{author}{\bibfnamefont{T.}~\bibnamefont{Mano}},
  \bibinfo{author}{\bibfnamefont{T.}~\bibnamefont{Noda}},
  \bibinfo{author}{\bibfnamefont{M.}~\bibnamefont{Yamagiwa}}, \bibnamefont{and}
  \bibinfo{author}{\bibfnamefont{N.}~\bibnamefont{Koguchi}},
  \bibinfo{journal}{unpublished}.

\bibitem[{\citenamefont{Kuroda et~al.}(2006)\citenamefont{Kuroda, Kuroda, Sakoda, Watanabe, Koguchi, Kido}}]{kuro06}
\bibinfo{author}{\bibfnamefont{K.}~\bibnamefont{Kuroda}},
  \bibinfo{author}{\bibfnamefont{T.}~\bibnamefont{Kuroda}},
  \bibinfo{author}{\bibfnamefont{K.}~\bibnamefont{Sakoda}},
  \bibinfo{author}{\bibfnamefont{K.}~\bibnamefont{Watanabe}},
  \bibinfo{author}{\bibfnamefont{N.}~\bibnamefont{Koguchi}}, \bibnamefont{and}
  \bibinfo{author}{\bibfnamefont{G.}~\bibnamefont{Kido}},
  \bibinfo{journal}{Appl. Phys.\ Lett.} \textbf{\bibinfo{volume}{88}},
  \bibinfo{pages}{124101} (\bibinfo{year}{2006}).
  
\bibitem[{\citenamefont{Pavesi and Guzzi}(1994)}]{pave94}
\bibinfo{author}{\bibfnamefont{L.}~\bibnamefont{Pavesi}} \bibnamefont{and}
  \bibinfo{author}{\bibfnamefont{M.}~\bibnamefont{Guzzi}},
  \bibinfo{journal}{J.\ Appl. Phys.} \textbf{\bibinfo{volume}{75}},
  \bibinfo{pages}{4779} (\bibinfo{year}{1994}).

\bibitem[{\citenamefont{Bosio et~al.}(1988)\citenamefont{Bosio, Staehli, Guzzi,
  Burri, and Logan}}]{bosi88}
\bibinfo{author}{\bibfnamefont{C.}~\bibnamefont{Bosio}},
  \bibinfo{author}{\bibfnamefont{J.~L.} \bibnamefont{Staehli}},
  \bibinfo{author}{\bibfnamefont{M.}~\bibnamefont{Guzzi}},
  \bibinfo{author}{\bibfnamefont{G.}~\bibnamefont{Burri}}, \bibnamefont{and}
  \bibinfo{author}{\bibfnamefont{R.~A.} \bibnamefont{Logan}},
  \bibinfo{journal}{Phys.\ Rev. B} \textbf{\bibinfo{volume}{38}},
  \bibinfo{pages}{3263} (\bibinfo{year}{1988}).
  \end{thebibliography}

\end{document}